\documentstyle[preprint,aps]{revtex}
\tightenlines
\begin{document}
\draft
\title{A POSSIBLE CRYPTO-SUPERCONDUCTING STRUCTURE IN A 
SUPERCONDUCTING FERROMAGNET}
\author{C. W. Chu$^{1,2}$, Y. Y. Xue$^{1}$, S. Tsui$^{1}$, 
J. Cmaidalka$^{3}$, A. K. Heilman$^{1}$, B. Lorenz$^{1}$\\ 
and R. L. Meng$^{1}$}
\address{$^{1}$Department of Physics and the Texas Center for Superconductivity, 
University of Houston, Houston, TX 77204-5932\\
$^{2}$Lawrence Berkeley National Laboratory, 1 Cyclotron Road, Berkeley, 
CA 94720\\
$^{3}$Department of Chemistry and Texas Center for Superconductivity, 
University of Houston, Houston, TX 77204-5932}

\date{\today}
\maketitle
\begin{abstract}
We have measured the $dc$ and $ac$ electrical and magnetic properties in various 
magnetic fields of the recently reported superconducting ferromagnet 
RuSr$_{2}$GdCu$_{2}$O$_{8}$. Our reversible magnetization measurements 
demonstrate the absence of a bulk Meissner state in the compound below the 
superconducting transition temperature. Several scenarios that might account 
for the absence of a bulk Meissner state, including the possible presence of a 
sponge-like non-uniform superconducting or a crypto-superconducting structure 
in the chemically uniform Ru-1212, have been proposed and discussed.
\end{abstract}
\pacs{\textit{Submitted to Physica C 
(November 1, 1999); accepted for 
publication (December 
24, 1999)}}

\section{INTRODUCTION}

It has long been demonstrated \cite{bulaevskii85} that, while an 
antiferromagnetic order can coexist with a uniform superconducting order (in a 
Meissner state), a long-range ferromagnetic order cannot. However, it was later 
shown \cite{bulaevskii85} that a non-uniform ferromagnetic order could coexist 
with superconductivity. This non-uniform ferromagnetic order manifests itself 
in the form of a spiral structure, or a domain-like structure, that reduces the 
compound's effectiveness in suppressing superconductivity. At the same time, 
it was also shown that superconductivity suppresses the ferromagnetic order. 
Many ternary compounds, comprised of both superconducting and ferromagnetic 
sublattices, indeed exhibit these fascinating magnetic structures in the 
superconducting state. For instance, ErRh$_{4}$B$_{4}$ \cite{sinha82} 
superconducts below its superconducting transition temperature ($T_{s}$) of 
8.7~K and orders ferromagnetically below its transition temperature 
($T_{m}$) of 0.8~K, before returning to a normal state below a re-entrant 
temperature ($T_{r}$) of 0.7~K. A spiral ferromagnetic structure was found to 
coexist with the superconducting state in this compound between $T_{m}$ and 
$T_{r}$. On the other hand, a domain-like ferromagnetic structure was observed 
between $T_{m}$ and $T_{r}$ of HoMo$_{6}$S$_{8}$ \cite{thomlinson82} and 
HoMo$_{6}$Se$_{8}$ \cite{lynn84} with ($T_{s}$, $T_{m}$, $T_{r}$) of 
(1.8~K, 0.7~K, 0.65~K) and (5.5~K, 0.53~K, 0~K), respectively. A spontaneous 
vortex lattice state was also proposed \cite{ferrel97} for the case in which 
the internal field of the ferromagnet ($4\pi M$) is greater than the lower 
critical field ($H_{c1}$) but lower than the upper critical field ($H_{c2}$). 
The compounds previously studied always have a $T_{s}$ higher than $T_{m}$ and 
have been known as ferromagnetic superconductors \cite{ferrel97}. In these 
ferromagnetic superconductors, superconductivity is the dominant state and the 
ferromagnetic state is modified with a non-uniform structure to fit the 
superconducting state.

Recently, several ruthenate-cuprates have been reported to undergo a transition 
at $T_{m}$ to a ferromagnetic state that coexists with a superconducting state 
at a lower transition temperature $T_{s}$. They are RuSr$_{2}$GdCu$_{2}$O$_{8}$ 
(Ru-1212) \cite{bernhard99} and 
RuSr$_{2}$(R$_{0.7}$Ce$_{0.3}$)$_{2}$Cu$_{2}$O$_{10}$ (Ru-1222) 
\cite{felner97}, in which R = Gd or Eu, with ($T_{m}$, $T_{s}$) of 
(133~K, 14--47~K) for Ru-1212, (180~K, 42~K) for Ru-1222 when R = Gd, and (122~K, 
32~K) for Ru-1222 when R = Eu. They have therefore been called superconducting 
ferromagnets \cite{sonin98}. The ferromagnetic state below $T_{m}$ has been 
firmly established \cite{felner97,sonin98} to be bulk and uniform to a scale 
of $\sim 20$~\AA\ across the samples, as examined by magnetization, 
muon-relaxation, and M\"ossbauer measurements. However, the cited evidence for 
superconductivity \cite{felner97,sonin98} rests entirely on the zero 
resistivity ($\rho$) observed and a diamagnetic shift of the $dc$ magnetic 
susceptibility ($\chi_{dc}$) measured in the zero-field-cooled (ZFC) mode. 
No diamagnetic shift has been detected in $\chi_{dc}$ of these samples when 
measured in the field-cooled (FC) mode. It is known that a diamagnetic shift in 
the ZFC-$\chi_{dc}$, which is a measure of the superconducting shielding effect 
and consistent with the zero-$\rho$ observed, is not proof for the existence of 
the Meissner state, a conventional criterion for the existence of bulk 
superconductivity. To determine the possibility of a nearly perfect pinning 
in Ru-1212 that reduces the size of the diamagnetic shift in FC-$\chi_{dc}$, we 
have recently determined \cite{chu99} the reversible magnetization ($M_{r}$) of 
the sample as a function of field ($H$) \cite{clem93}. The results showed that a 
Meissner state exists in no more than a few percent of the bulk sample, if it 
exists at all. Several possibilities have been proposed \cite{chu99} to account 
for the absence of a bulk Meissner state in Ru-1212: the possible formation of a 
spontaneous vortex state in the bulk superconducting ferromagnet Ru-1212; the 
possible presence of a spatially non-uniform superconducting structure in the 
chemically uniform Ru-1212; and the possible appearance of a filamentary 
superconductivity associated with a crypto-superconducting fine structure in a 
chemically uniform superconducting ferromagnet or with an impurity phase in 
this otherwise non-superconducting ferromagnet. In this paper, we further 
examine these possibilities proposed for the absence of the bulk Meissner 
effect in superconducting ferromagnets. 

\section{EXPERIMENTAL}

Ru-1212 samples investigated here were either the same ones previously 
studied \cite{chu99} or freshly prepared by thoroughly reacting a mixture of 
RuO$_{2}$, SrCO$_{3}$, Gd$_{2}$O$_{3}$, and CuO, with a cation composition of 
Ru:Sr:Gd:Cu = 1:2:1:2, following steps previously reported, including the 
prolonged annealing at 1060~$^{\circ}$C 
\cite{bernhard99,felner97,chu99}. The structure was determined by powder X-ray 
diffraction (XRD), using the Rigaku DMAX-IIIB diffractometer; the resistivity 
by the standard four-lead technique, employing the Linear Research Model LR-700 
Bridge; and both the $dc$ and $ac$ magnetizations by the Quantum Design SQUID 
magnetometer. Powder samples were prepared by pulverizing the pellets after 
prolonged annealing and selecting particles of different sizes by using sieves 
of different meshes.

\section{RESULTS AND DISCUSSION}

The samples investigated are pure Ru-1212 within the XRD resolution of a few 
percent. They display a tetragonal structure with a space group $P4/mmm$. The 
lattice parameters are $a = 3.8375(8)$~\AA\ and $c = 11.560(2)$~\AA\, in good 
agreement with values reported previously \cite{bernhard99,chu99}. The scanning 
electron microscope results show grains of 1--5~$\mu$m in size and the scanning 
microprobe data show a uniform composition across the sample to a scale 
of 1--2~$\mu$m.

The temperature dependence of $\chi_{dc}$ is shown in Fig.~\ref{cvt} and was 
measured at 5~Oe during both the ZFC and FC modes. A strong diamagnetic shift 
at $\sim 25$~K is detected in the ZFC-$\chi_{dc}$, but not in the FC-$\chi_{dc}$, 
similar to previous studies \cite{bernhard99,chu99}. It should be noted that 
such a diamagnetic shift in ZFC-$\chi_{dc}$ was sometimes absent from some 
samples that have the same XRD pattern. The temperature dependence 
of $\rho$ is exhibited in Fig.~\ref{rvt}. A drastic $\rho$-drop starting at 
$\sim 45$~K and reaching zero at $\sim 30$~K ($\equiv T_{s}$) is observed. In 
the presence of a magnetic field, $T_{s}$ is suppressed to $\sim 8$~K at 7~T, 
characteristic of a superconducting transition below $T_{s}$.

To determine if the absence of a diamagnetic shift in the FC-$\chi$ in the 
Ru-1212 sample is associated with a possible strong pinning, we determined 
$M_{r}$ \cite{clem93} as a function of $H$ from negative to positive $H$ in 
very fine $H$-steps, especially in the low-$H$ region. Since the surface pinning 
in cuprate superconductors is small, $M_{r}$ can be taken as 
$(1/2)(M_{+} + M_{-})$, where $M_{+}$ and $M_{-}$ are magnetizations measured 
during the increasing and the decreasing of the field, respectively. The results 
of $M_{r}(H)$ and $dM_{r}(H)/dH$ are summarized in Fig.~\ref{mvh}. $M_{r}$ does 
not exhibit any clear deviation from the almost linear $M_{r}$-$H$ relation; nor 
does $dM_{r}(H)/dH$ display the large initial increase expected of a Meissner 
state as $H$ passes over $H_{c1}$, even for $H$ as small as $\pm 0.25$~Oe. The 
results demonstrate unambiguously the absence of a bulk Meissner state in the 
Ru-1212 samples investigated down to 0~Oe. The slight initial increase in 
$dM_{r}/dH$ with $H$ enables us to estimate that no more than a few percent of 
the bulk sample is in the true Meissner state, if the state exists at all.

In a superconducting ferromagnet with a $T_{m} > T_{s}$, a spontaneous 
magnetization $M$ occurs below $T_{m}$. If the internal field associated with 
the moments $4\pi M$ is greater than $H_{c1}$ but smaller than $H_{c2}$, a 
spontaneous vortex lattice has been proposed to form below $T_{s}$, similar to 
that predicted \cite{ferrel97} for a ferromagnetic superconductor where 
$T_{s} > T_{m}$. Since $4\pi M$ and $H_{c1}$ are temperature-dependent, there 
may exist two scenarios for the superconducting ferromagnet. When its $4\pi M$ 
is always greater than $H_{c1}$ below $T_{m}$, the compound will undergo a 
paramagnetic-to-ferromagnetic transition at $T_{m}$ and then enter the 
spontaneous vortex state at and below $T_{s}$, as shown in 
Fig.~\ref{formation}a. This was proposed to be a possible case for Ru-1212 
\cite{chu99}. When the growth of $H_{c1}$ outpaces that of $4\pi M$ below a 
certain temperature $T_{so}$, as shown in Fig.~\ref{formation}b, the compound 
first undergoes a paramagnetic-to-ferromagnetic transition at $T_{m}$ upon 
cooling and then enters the spontaneous vortex state at $T_{s}$, before it 
recovers the Meissner state at $T_{so}$. Such a case has been suggested to take 
place in Ru-1222 \cite{sonin98}, although the Meissner state below $T_{so}$ was 
not evident in their FC-$\chi_{dc}$. Another possible scenario is a triplet 
superconducting pairing, in which $T_{m}$ may coincide with $T_{s}$ 
\cite{knigavko98}. In this scenario, the compound will directly enter the 
spontaneous state at $T_{s}$ from its paramagnetic state when $4\pi M > H_{c1}$ 
(Fig.~\ref{formation}c) or recover its Meissner state if $4\pi M$ becomes 
smaller than $H_{c1}$ below $T_{so}$ (Fig.~\ref{formation}d). We conjecture 
that the superconducting Sr$_{2}$RuO$_{4}$ may fall into this third category. 

In the above discussions, a bulk superconducting state is assumed to exist in 
the superconducting ferromagnet below $T_{s}$. Its existence is usually best 
demonstrated by the detection of a specific heat anomaly ($\Delta C_{p}$) at 
$T_{s}$. A $\Delta C_{p}$ was indeed reported \cite{tallon} in Ru-1212 near 
$T_{s}$, with a magnitude close to that of an underdoped superconducting 
cuprate. Unfortunately, it was also reported that while the magnitude of 
$\Delta C_{p}$ is suppressed by an externally applied $H$, as expected of a 
bulk superconducting transition, the peak temperature of $\Delta C_{p}$ 
increases with $H$, in strong contrast to the $T_{s}$-suppression by $H$ 
observed. This suggests that the detected $\Delta C_{p}$ may be of a magnetic 
origin. It is known \cite{chu99} that Sr$_{2}$GdRuO$_{6}$, which forms easily 
in the Ru-1212 matrix, is an antiferromagnet with a Neel temperature close to 
$T_{s}$ at $\sim 30$~K and a transition drastically suppressed by $H$ 
\cite{chu99}. The presence of a few tenths of a percent of Sr$_{2}$GdRuO$_{6}$ 
as an impurity in Ru-1212 can easily account for the $\Delta C_{p}$-anomaly 
detected near $T_{s}$ \cite{tallon}.

In view of the possible absence of $\Delta C_{p}$, and thus of bulk 
superconductivity in Ru-1212, we decided to investigate both the chemical and 
the electrical uniformity of the Ru-1212 samples. As mentioned earlier, the XRD 
pattern of our samples shows only the Ru-1212 phase within our experimental 
resolution of a few percent. The microprobe scan indicates that the sample is 
chemically uniform to a scale of $\sim$~1--2~$\mu$m in agreement with previous 
report \cite{bernhard99}. To determine the electrical uniformity of the sample 
in terms of its superconducting property, we examined the superconducting 
shielding effect of Ru-1212 powder samples of different particle sizes by 
measuring the $ac$ magnetization ($M_{ac}$) as a function of the $ac$ field 
($H_{ac}$) in fine steps to $\sim 10^{-4}$~Oe in zero $dc$ field. Details of the 
experiment will be published elsewhere \cite{xue}. Typical results of the $ac$ 
susceptibility $\chi_{ac} \equiv M_{ac}/H_{ac}$ for samples of particle sizes 
down to 20~$\mu$m, all pulverized sequentially from the same bulk sample of 
$\sim 6$~mm, at 2~K are displayed in Fig.~\ref{cvh}. For the 6~mm bulk sample, 
$-4\pi \chi_{ac}$ varies negligibly with $H_{ac} < 0.1$~Oe, but decreases 
rapidly with $H_{ac} > 1$~Oe. The $ac$ superconducting shielding effect as 
measured by $-4\pi \chi_{ac}$ before the demagnetization factor correction is 
$\sim 200$\%. As the particle size decreases, $-4\pi \chi_{ac}$ decreases, but 
remains almost $H_{ac}$-independent until $H_{ac} > 1$~Oe. To assure that the 
$-4\pi \chi_{ac}$-drop is not caused by the degradation of the samples due to 
pulverization, we compared the $M_{r}$'s for both the bulk and the powdered 
sample of a particle-size of 50~$\mu$m as shown in the inset in 
Fig.~\ref{cvt}. The same $M_{r}$ was detected for the two samples, suggesting 
that the powdered samples retain their integrity. For a uniform bulk 
superconductor, $\chi_{ac}$ is expected to be independent of the particle-size, 
provided that the penetration depth $\lambda \ll d$, where $d$ is the size of the 
particle. However, the field $H'_{ac}$, beyond which $-4\pi \chi_{ac}$ starts to 
decrease, is determined by $J_{c}d$, where $J_{c}$ is the critical current 
density of the compound, and is expected to decrease with $d$, according to the 
Bean model. This is borne out by our results for a similar experiment on the 
uniform bulk and powdered superconducting YBa$_{2}$Cu$_{3}$O$_{7}$ 
\cite{xue}. Therefore, our observation in Ru-1212 is in variance with what is 
expected of a uniform bulk superconductor in two respects: the 
$\chi_{ac}$-decrease with $d$ and the almost $d$-independent $H'_{ac} 
\sim 0.5$~Oe. The $H'_{ac}$ predicted by the Bean model is also shown in 
Fig.~\ref{cvh} for comparison. These variances can be explained by assuming 
that Ru-1212 is a bulk superconductor and has a penetration depth 
$\lambda \sim$~50~$\mu$m. However, the estimated $\lambda \sim$~50~$\mu$m seems 
to be too large for a cuprate superconductor which has a reported 
$\lambda$ value usually about 20--100 times smaller. To explain the large 
$\lambda$, we assume that there may exist a sponge-like superconducting network 
uniformly distributed across the Ru-1212 sample to a scale of $< 20$~$\mu$m. 
In other words, the superconducting order may have modified itself with 
such a non-uniform fine structure (crypto-superconductivity) in order to fit 
into the ferromagnetic state in the superconducting ferromagnet Ru-1212, in a 
way very similar to the proposed crypto-ferromagnetism in a ferromagnetic 
superconductor \cite{anderson59}. For instance, superconductivity may exist 
between the ferromagnet-domain boundaries where the magnetic field can be 
smaller than $H_{c2}$ and the magnetic scattering is suppressed. The formation 
of these fine ferromagnet-domains in the chemically uniform itinerant 
ferromagnet Ru-1212, driven by the electromagnetic interaction, appears to be 
not unreasonable in view of what has been observed in the closely related 
highly correlated colossal magnetoresistance manganites 
\cite{goodenough97}. Such a superconducting structure is expected to display a 
very small condensation energy associated with the superconducting transition 
at $T_{s}$. This proposed structure is consistent with the zero $\rho$ observed 
and with our preliminary results suggesting a negligible condensation energy 
determined by the magnetization method \cite{xue}.

While the large $\lambda$ suggested by our $\chi_{ac}$-$H_{ac}$ results can be 
explained in terms of the possible appearance of crypto-superconductivity, we 
would also like to address the other possible cause, \textit{i.e.} the presence 
of a minor superconducting impurity phase in the otherwise non-superconducting 
ferromagnet Ru-1212. RSr$_{2}$Cu$_{3}$O$_{7}$ (Cu-1212) with R = Y or a 
rare-earth element is known to form at ambient pressure by slight doping with a 
high valence element \cite{greaves89} or under high pressure without doping 
\cite{okai90}. For example, the Cu-1212 with its chain-Cu partially replaced by 
Ru has been reported \cite{chen95} to form at ambient pressure and has a 
$T_{s}$ of $\sim$~20--65~K. A small amount of the slightly Ru-doped Cu-1212 
present in an otherwise non-superconducting Ru-1212 may thus account for the 
observation of the large $\lambda$, the zero $\rho$, and the absence of a bulk 
Meissner state. However, more than $\sim 10$\% of the minor phase of Ru-doped 
Cu-1212 is required to be present in order to account for the observed 
percolative path of zero $\rho$ in Ru-1212. Although the similar structures of 
Ru-1212 and Cu-1212 make the XRD less decisive in differentiating the two 
phases, the difference between the Cu/Ru-ratios of the two phases (2/1 and 
2.7/0.3 for the Ru-1212 and Ru-doped Cu-1212, respectively) can be easily 
detected by the microprobe for their separate presence. Unless the slightly 
Ru-doped Cu-1212 or other unknown superconducting impurity phase is finely 
dispersed to a scale below one micrometer or is deposited in the grain 
boundaries of the sample, our microprobe data suggest the unlikely presence of 
such superconducting impurities, in agreement with previous reports 
\cite{bernhard99}.

In conclusion, we have measured the electrical and magnetic properties of both 
bulk and powdered Ru-1212 of different particle sizes in various $dc$ and $ac$ 
fields. The XRD and microprobe data show that the Ru-1212 samples examined are 
chemically pure to a few percent and uniform to a scale of 1--2~$\mu$m. The 
reversible magnetization as a function of field clearly demonstrates the 
absence of a bulk Meissner state in Ru-1212. The $-4\pi \chi_{ac}$ decreases as 
the particle size decreases and remains constant initially as the $ac$ field 
increases, but decreases rapidly, independent of the particle size, as the $ac$ 
field increases to above $\sim 0.5$~Oe. Several possible scenarios have been 
considered and discussed. While a spontaneous vortex state may exist in a bulk 
superconducting ferromagnet Ru-1212, we also propose that the superconducting 
state in a superconducting ferromagnet is more likely to have modified itself 
to a sponge-like non-uniform superconducting fine (crypto-superconducting) 
structure to fit the ferromagnetic state in the chemically uniform Ru-1212. 
While the spontaneous vortex lattice shows a reduced condensation energy as the 
compound undergoes the superconducting transition, the crypto-superconducting 
structure is expected to display a negligible condensation energy. Further 
experiments to determine the condensation energy and the spontaneous vortex 
lattice in superconducting ferromagnets are needed to differentiate the above 
suggestions.

\acknowledgments

The work in Houston is supported in part by NSF Grant No.~DMR-9804325, 
the T.~L.~L. Temple Foundation, the John~J.~and Rebecca Moores Endowment, and 
the State of Texas through the Texas Center for Superconductivity at the 
University of Houston; and at Lawrence Berkeley Laboratory by the Director, 
Office of Energy Research, Office of Basic Energy Sciences, Division of 
Materials Sciences of the U.S. Department of Energy under Contract 
No.~DE-AC03-76SF00098.

\begin{figure}
\caption{$4\pi \chi_{dc}$ \textit{vs} $T$ for Ru-1212 at 5~Oe. 
Inset: $M_{r}$ \textit{vs} $H$ at 2~K for Ru-1212 bulk ($\bigtriangledown$) and 
50~$\mu$m particles ($\bigcirc$).}
\label{cvt}
\end{figure}

\begin{figure}
\caption{$\rho$ \textit{vs} $T$ for Ru-1212 at 0 (solid) and 9 T (dashed).}
\label{rvt}
\end{figure}

\begin{figure}
\caption{$M_{r}$ and $dM_{r}/dH$ \textit{vs} $H$ for Ru-1212 at 2~K.}
\label{mvh}
\end{figure}

\begin{figure}
\caption{The formation of the spontaneous vortex state (SVS) and Meissner state 
(MS) in superconducting ferromagnets, in which $T_{m} > T_{s}$, when 
a) $H_{c1}$ is always smaller than $4\pi M$; b) $H_{c1} > H_{c1}$ below $T_{so}$; 
c) $T_{s} = T_{m}$ and $H_{c1} < 4\pi M$ always; and d) $T_{s} = T_{m}$ and 
$H_{c1} > 4\pi M$.}
\label{formation}
\end{figure}

\begin{figure}
\caption{$\chi_{ac}$ \textit{vs} $H_{ac}$ for bulk and powdered samples of 
different particle sizes: $\bigtriangleup$ --- bulk; $\Box$ --- 820~$\mu$m; 
$\bigtriangledown$ --- 110~$\mu$m; $\Diamond$ --- 50~$\mu$m; and 
$\bigcirc$ --- 
20~$\mu$m. The arrows represent the $H'_{ac}$ based on the Bean model.}
\label{cvh}
\end{figure}


\begin{references}
\bibitem{bulaevskii85}For a review, see L. N. Bulaevskii, A. I. Buzdin, 
M. L. Kuli\'c and S. V. Panjukov, Advances in Physics 34, 175 (1985).
\bibitem{sinha82}S. K. Sinha, H. A. Mook, D. G. Hinks and G. W. Crabtree, 
Phys. Rev. Lett. 48, 950 (1982).
\bibitem{thomlinson82}W. Thomlinson, G. Shirane, J. W. Lynn and D. E. Moncton, 
Superconductivity in Ternary Compounds II, Topics in Current Physics, vol. 34, 
ed. M. B. Maple and \O. Fischer (Berlin: Springer-Verlag, 1982), p. 99.
\bibitem{lynn84}J. W. Lynn, J. A. Gotaas, R. W. Erwin, R. A. Ferrel, J. K. 
Bhattacharjee, R. N. Shelton and P. Klevins, Phys. Rev. Lett. 52, 133 (1984).
\bibitem{ferrel97}R. A. Ferrel, J. K. Bhattacharjee and A. Bagchi, Phys. Rev. 
Lett. 43, 154 (1997); H. Matsumota, H. Umezawa and M. Tachiki, S. S. Comm. 31, 
157 (1979); and H. S. Greenside, E. I. Blount and C. M. Varma, Phys. Rev. Lett. 
46, 49 (1981).
\bibitem{bernhard99}C. Bernhard, J. L. Tallon, Ch. Niedermayer, Th. Blasillo, A. 
Golnik, E. BrŸcker, K. K. Kremer, D. R. Noakes, C. E. Stronack and E. J. 
Ansaldo, Phys. Rev. B 59, 14099 (1999).
\bibitem{felner97}I. Felner, U. Asaf, Y. Levi and O. Millo, Phys. Rev. B 55, 
R3374 (1997).
\bibitem{sonin98}E. B. Sonin and I. Felner, Phys. Rev. B 57, R14000 (1998).
\bibitem{chu99}C. W. Chu, Y. Y. Xue, R. L. Meng, J. Cmaidalka, L. M. Dezaneti, 
Y. S. Wang, B. Lorenz and A. K. Heilman, cond-mat/9910056 5 October 1999.
\bibitem{clem93}J. R. Clem and Z. D. Hao, Phys. Rev. B 48, 13774 (1993).
\bibitem{knigavko98}A. Knigavko and B. Rosenstein, Phys. Rev. B58, 9354 (1998).
\bibitem{tallon}J. L. Tallon, J. W. Loram, G. V. M. Williams and C. Bernhard, 
to appear in Phys. Rev. Lett.
\bibitem{xue}Y. Y. Xue, S. Tsuei, J. Cmaidalka and C. W. Chu, to be published.
\bibitem{anderson59}P. W. Anderson and H. Suhl, Phys. Rev. 116, 898 (1959).
\bibitem{goodenough97}J. Goodenough and J. S. Zhou, Nature 386, 229 (1997).
\bibitem{greaves89}C. Greaves and P. R. Slater, Physica C 180, 245 (1989); 
T. E. Dann, M. S. Thesis, Tankung University, Taiwan (1990); and B. Debrowski, 
K. Rogack, J. W. Koenitzer, K. R. Peoppelmeir and J. D. Jorgensen, Physica C 
277, 24 (1997).
\bibitem{okai90}B. Okai, Jpn. J. Appl. Phys. 29, L2180 (1990); and Y. Cao, 
T. L. Hudson, Y. S. Wang, J. H. Xu, Y. Y. Xue and C. W. Chu, Phys. Rev. B 58, 
11201 (1998).
\bibitem{chen95}D. H. Chen, Ph.D. Thesis, National Tsing Hua University, 
Taiwan (1995).
\end{references}
\end{document}